%% file: main.tex
\documentclass{article} 
\usepackage{iclr2026_conference,times}

\input{math_commands.tex}

\usepackage{hyperref}
\usepackage{url}
\usepackage{graphicx}
\usepackage{wrapfig}
\usepackage{float}
\usepackage{enumitem}

\usepackage{amsmath}
\usepackage{amssymb}
\usepackage{mathtools}
\usepackage{multirow}
\usepackage{multicol} 
\usepackage{booktabs}
\usepackage{amsthm}
\usepackage{adjustbox}

\usepackage{subfigure} 
\usepackage{setspace}

\usepackage{comment}

\usepackage[noabbrev,capitalize]{cleveref}

\crefname{equation}{equation}{equations}
\crefname{line}{line}{lines}
\crefname{section}{\S}{\S\S}
\Crefformat{section}{#2\S#1#3}
\crefrangeformat{section}{\S\S#3#1#4--#5#2#6}

\newcommand{\modelname}{\textsc{Sigma-MoE}}
\newcommand{\project}{\textsc{Sigma}}
\newcommand{\ltp}{\textsc{Lucia Training Platform}}
\newcommand{\ltf}{\textsc{Lucia Training Framework}}

\title{\project{}: An AI-Empowered Training Stack \\ on Early-Life Hardware}

\iclrfinalcopy
\author{\project{} v-Team\\Microsoft Research}

\begin{document}

\maketitle

\input{sections/abstract}
\input{sections/overview}
\input{sections/challenges}
\input{sections/platform}
\input{sections/framework.tex}
\input{sections/model}

\input{sections/results}

\input{sections/conclusion}

\clearpage

\bibliography{iclr2026_conference}
\bibliographystyle{iclr2026_conference}

\appendix
\input{sections/appendix}

\end{document}

%% file: math_commands.tex
\usepackage{amsmath,amsfonts,bm}

\def\eqref#1{equation~\ref{#1}}

\def\1{\bm{1}}

\DeclareMathAlphabet{\mathsfit}{\encodingdefault}{\sfdefault}{m}{sl}
\SetMathAlphabet{\mathsfit}{bold}{\encodingdefault}{\sfdefault}{bx}{n}



%% file: sections/abstract.tex
\begin{abstract}
An increasing variety of AI accelerators is being considered for large-scale training. However, enabling large-scale training on early-life AI accelerators faces three core challenges: frequent system disruptions and undefined failure modes that undermine reliability; numerical errors and training instabilities that threaten correctness and convergence; and the complexity of parallelism optimization combined with unpredictable local noise that degrades efficiency.
To address these challenges, \project{} is an open-source training stack designed to improve the reliability, stability, and efficiency of large-scale distributed training on early-life AI hardware.
The core of this initiative is the \ltp{} (LTP), the system optimized for clusters with early-life AI accelerators. Since its launch in March 2025, LTP has significantly enhanced training reliability and operational productivity. Over the past five months, it has achieved an impressive 94.45\% effective cluster accelerator utilization, while also substantially reducing node recycling and job-recovery times.
Building on the foundation of LTP, the \ltf{} (LTF) successfully trained \modelname{}, a 200B MoE model, using 2,048 AI accelerators. This effort delivered remarkable stability and efficiency outcomes, achieving 21.08\% MFU, state-of-the-art downstream accuracy, and encountering only one stability incident over a 75-day period.
Together, these advances establish \project{}, which not only tackles the critical challenges of large-scale training but also establishes a new benchmark for AI infrastructure and platform innovation, offering a robust, cost-effective alternative to prevailing established accelerator stacks and significantly advancing AI capabilities and scalability. The source code of \project{} is available at \url{https://github.com/microsoft/LuciaTrainingPlatform}.\footnote{The paper results from an open source project.}

\end{abstract}

%% file: sections/overview.tex
\section{Overview}
In the wake of the large language model boom, a diverse range of AI hardware is being explored for large-scale training. However, utilizing these early-life accelerators for large-scale training presents three main challenges.
First, \emph{reliability}: frequent system disruptions and poorly characterized failure modes in the underlying infrastructure layer increase operational cost and erode user confidence.
Second, \emph{stability}: subtle numerical errors and algorithmic instabilities are hard to detect and remediate in the platform or framework layer, threatening correctness and convergence.
Third, \emph{efficiency}: achieving and maintaining high throughput in the framework and application layers requires continual parallelism tuning because device efficiencies and training dynamics vary, and local disruptions can produce large, hard-to-predict fluctuations in overall performance.

Existing training software stacks are ill-equipped to address these challenges on early-life AI hardware.
We therefore introduce \project{}, a pioneering open-source training software stack designed to enable large-scale distributed training on early-life AI hardware once vendor libraries become available.
\project{} collapses the conventional multi-layer AI software stack into two cleanly separated layers, a unified platform layer (\ltp{}) and a scalable framework layer (\ltf{}), so that hardware reliability and fault mitigation are handled centrally in the platform while the framework focuses on numerical stability and training efficiency.
By integrating advanced learning-based AI techniques across the stack, \project{} systematically improves reliability, stability, and efficiency of large-scale distributed training, thereby substantially improving user experience and operational productivity.

\project{}'s platform layer, \ltp{} (LTP), showcases AI-driven, continuous improvements in large-scale training reliability on early-life hardware.
The platform pioneers an end-to-end, AI-driven innovation pipeline.
Since April 2025, LTP has been deployed on a cluster of 2,144 AI accelerators; by July 2025 it sustained uninterrupted training runs exceeding 79 hours on 2,048-accelerator jobs, marking a significant
breakthrough in large-scale reliability.
Key achievements include a $94.45\%$ effective utilization (with an additional $3.37\%$ reserved for backups), corresponding to a $7.90\times$ improvement over the initial cluster baseline.
Cluster efficiency improved dramatically: node recycling time fell from 116.36 hours to 11.87 hours, and average job recovery time dropped from 2.5 hours to under 10 minutes.
These gains increase training reliability and operational efficiency while yielding substantial cost savings, positioning LTP as a viable, competitive alternative to prevailing established accelerator infrastructure and platforms.

\project{}'s framework layer, \ltf{} (LTF), demonstrates stable and efficient training on early-life hardware through a large-scale training of \modelname{}, a 200B-parameter MoE model with 20B activated parameters, on 2,048 AI accelerators.
During a 75-day training run we observed a stability issue only \textit{once}, which was promptly resolved by improving the MoE load-balancing loss function. In terms of efficiency, our setup achieved $21.08\%$ Model FLOPS Utilization (MFU). The impressive results of \modelname{}, comparable with DeepSeek V2, demonstrate the success of \project{}'s end-to-end pipeline.

Through this initiative, \project{} not only simplifies the AI software stack but also directly addresses the core challenges of reliability, stability, and efficiency in large-scale training on early-life AI hardware by integrating advanced learning-based techniques across the stack.
The development of LTP empirically demonstrates these gains, achieving substantial improvements in availability and reliability on early-life AI accelerator clusters, and provides a solid foundation for large-scale training.
Built atop this foundation, the successful training of \modelname{} with LTF shows that impressive accuracy, stability, and efficiency are attainable even under constrained resources.
Together, these contributions yield \project{}, an open-source training stack that facilitates transformative advances in AI capabilities and fosters the co-evolution of AI models and AI clusters, driving significant improvements in operational productivity and user experience.

Note that while the platform supports various models, this paper's scope is strictly limited to the framework and platform, and does not include an assessment or review of the specific models.

%% file: sections/challenges.tex
\section{Challenges for Large-Scale Training on Early-Life Hardware}

\subsection{Reliability}

The rapid increase of large language models (LLMs) has placed unprecedented demands on the availability and reliability of AI computing clusters, especially on large-scale distributed training jobs that span thousands of AI accelerators due to the synchronous nature of training.
However, the deployment of early-life AI accelerator platforms introduces significant reliability challenges that can severely impact operational efficiency, user experience and significant time and cost overheads.

\paragraph{Prevalent Frequent System Disruptions.}

\begin{figure*}[t]
    \centering
    \hspace{0.8\textwidth}
    \includegraphics[width=0.7\textwidth]{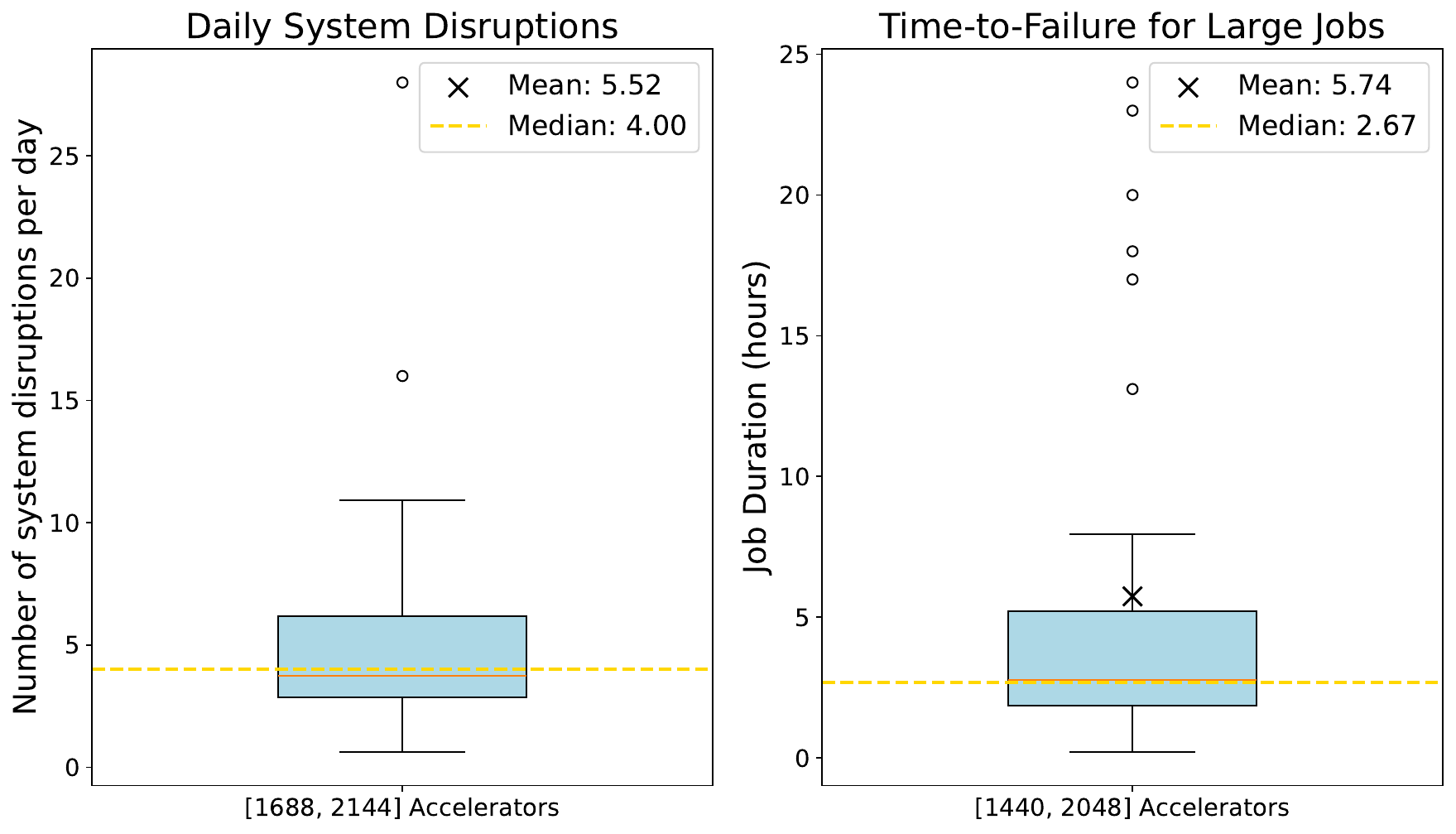}
    \caption{Distribution of Daily System Disruptions Counts and Time-to-Failure for Large Jobs on Early-Life AI Accelerators.}
    \label{fig:failure}
\end{figure*}

While Large-scale training on early-life accelerators faces a cascade of reliability challenges, fundamentally shaped by frequent system disruptions and their systemic impact. In a production environment exceeding 1,600 AI accelerators, we observed an average of 5.52 system disruptions per day during the initial deployment phase as shown in the left of Figure \ref{fig:failure}.
Due to the synchronous nature of distributed LLM pretraining, a single node failure can interrupt the entire job, causing the job-level Mean Time Between Failures (MTBF) to decrease sharply as scale increases. For example, as depicted on the right side of Figure \ref{fig:failure}, the median MTBF for a job involving over 1,440 AI accelerators was as low as 2.67 hours, significantly compromising operational stability.

The frequent occurrence of system disruptions on early-life AI accelerators directly diminishes overall resource availability. Each disrupted node necessitates a multi-step node remediation process: affected nodes need to be diagnosed and even go through the repair often spanning several days. As a result, a persistent fraction of the cluster remains unavailable, reducing effective compute capacity and inflating operational costs. The chronic gap between nominal and effective capacity translates into thousands of wasted accelerator-hours weekly, directly impacting both research velocity and financial efficiency.

Prolonged job downtime further exacerbates resource waste and demands for high platform productivity. A single 6-hour recovery for a 2,048-accelerator job can result in over 12,000 idle accelerator-hours. According to the MTBF, a job may run successfully for only a few hours before failing, after which the user must wait for a day or more for the faulted node to be restored. This dynamic drastically inflates the end-to-end training duration and, more importantly, erodes user confidence in the reliability of the computing infrastructure.
To mitigate this, the system must support rapid fault detection and recovery, necessitating more frequent and efficient DRI (Designated Responsible Individual) coverage, which increases operational overhead. 
This also place a significant cognitive and operational burden on users. Instead of focusing on model development, users are compelled to engage in failure diagnosis, and coordination with platform operators, detracting from research productivity.

\paragraph{Undefined and Continuously Emerging System Disruptions Types.}

A key reliability consideration for platform on early-life accelerators is the diverse and evolving nature of system disruptions. Unlike mature platform, which already have standardized fault logging and monitoring interfaces, the system platform on early-life accelerators evolve rapidly, leading to new and not yet fully documented disruption patterns. 

\begin{figure}[t]
    \centering
    \includegraphics[width=0.99\textwidth]{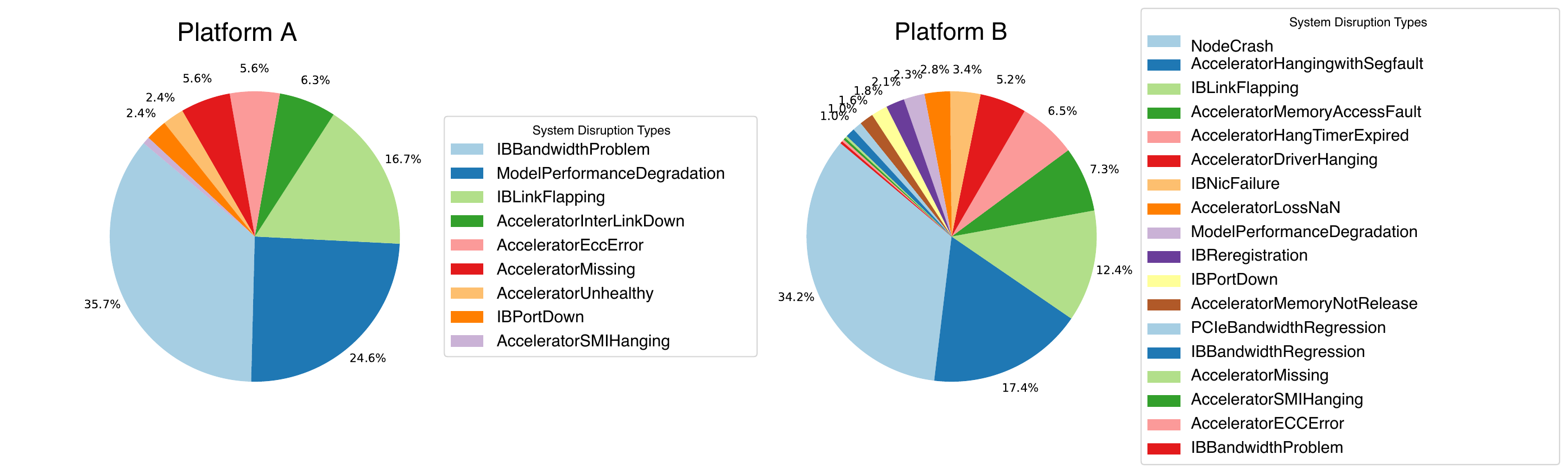}
    \caption{Failure Type Distribution of Different AI Platforms}
    \label{fig:reliability-challenges}
\end{figure}

As illustrated in Figure~\ref{fig:reliability-challenges}, our observations show a contrast between different platforms which includes many different types of accelerators from different vendors: while one platform exhibits clearly defined disruption modes, another shows a range of different and newly undefined system disruption types. During our operational observation period, a notable proportion of system disruptions were categorized as undefined or different from mature platform. While many disruptions on mature platforms can already be promptly classified and addressed, certain events on early-life platforms may lack clearly defined error codes or actionable telemetry, often manifesting as ambiguous system errors or silent hangs. This ambiguity makes fault detection and diagnosis far more difficult and time-consuming, often requiring expert intervention, manual log analysis and metrics correlation analysis.

The absence of defined system disruption categorization and monitoring standards not only impedes automated detection, but also delays the development of effective recovery workflows. Existing alerting systems, designed for mature platforms, frequently miss these new failures. 

\subsection{Stability}
\label{subsec:challenge-stability}

\begin{table}[!t]
\centering
\caption{Selected Stability Failures for Training on Early-Life AI Accelerator}
\begin{tabular}{@{}cccccc@{}}
\toprule
Failure & Diagnostic Time & Wasted Training Time & Root Cause  \\ \midrule
A & 4 days & 0.03 days & Numerical bug in flash-attention \\
B & 11 days & 0.4 days & Improper activation ratio and weight decay \\
C & 14 days & 15 days & Improper FFN initialization \\ \bottomrule
\end{tabular}
\label{tab:stability-failures}
\end{table}

Maintaining training stability is essential for ensuring efficient end-to-end model training within a reasonable time frame. Achieving consistent convergence of the optimization loss requires the correct implementation of both the training infrastructure—comprising the software stack and hardware—and the model training recipe, which includes the architecture, algorithm, and data. During our exploration of \modelname{} training on on early-life AI accelerator platforms, we encountered several stability failures, as detailed in \cref{tab:stability-failures}. These issues not only waste resources because the training efforts from the healthy checkpoint to the point of failure become futile, but also lead to prolonged root cause analysis, fundamentally delaying the overall training process. Below, we summarize the two key challenges inherent in maintaining training stability:

\paragraph{Emerging Hardware Complexity.} Training on early-life hardware platforms introduces additional complexity, particularly in terms of stability. New software stacks and hardware configurations raise the likelihood of numerical errors. For instance, \cref{tab:module-errors} outlines the unique numerical errors we observed in the software stack for early-life AI accelerators during our 4-month training experience.
Furthermore, the higher frequency of these errors prolongs the overall diagnostic time, as it becomes challenging to determine whether the issue arises from the infrastructure or the training recipe. This necessitates rigorous validation and testing to ensure correctness.

\begin{table}[!t]
\centering
\caption{Numerical Errors Example Identified in Software Stack on Early-Life AI Accelerator.}
\begin{tabular}{@{}cccccc@{}}
\toprule
Module & Error Details  \\ \midrule
Flash-attention & Error in gradient computation with vendor-specific backend \\
BLAS & Diverged results in GEMM for equal tiles \\
Grouped GEMM & Error in some ranks with expert parallelism enabled \\ 
Layer Norm & Accumulated gradient error in attention k-layernorm \\ \bottomrule
\end{tabular}
\label{tab:module-errors}
\end{table}

\begin{figure}[htbp]
  \centering
  \includegraphics[width=0.48\textwidth]{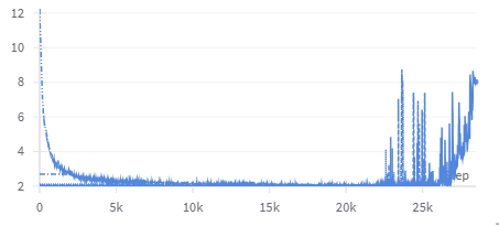}
  \includegraphics[width=0.48\textwidth]{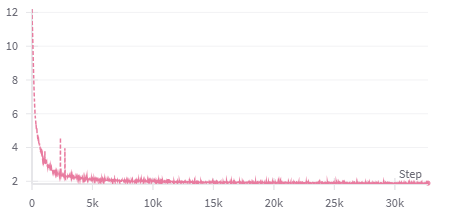}
  \caption{Per-step loss curves of an early version of \project{}-V2 model. The left curve is before fixing the model and training recipe design issue, and the right one is after the fix.}
  \label{fig:stability-loss-curve}
\end{figure}

\paragraph{Delayed Training Collapse Detection.} Even if the underlying software stack is numerically correct, improper model and training recipe design can lead to stability failures, as seen in Events B and C in \cref{tab:stability-failures}. Instabilities can silently accumulate during the model training process, remaining undetected for tens of thousands of iterations until the loss curve abruptly diverges, resulting in significant wasted training progress. \cref{fig:stability-loss-curve} illustrates a typical example where instability is only detected after 25K training steps. This issue stems from flaws in the model and training recipe design for large models and does not occur with smaller models. Therefore, early and accurate collapse detection methods are needed to halt the faulty training process well before a complete collapse occurs, minimizing the waste of training progress.

As a result, ensuring the correctness of the training infrastructure on early-life hardware and prompt training collapse detection are crucial for ensuring stable and efficient end-to-end model training.

\subsection{Efficiency}
\label{subsec:challenge-efficiency}
Efficiency is another critical factor influencing the completion time of training workloads. The two key challenges are as follows:

\paragraph{Frequent Parallelism Tuning.} Optimizing model parallelism is crucial for enhancing training efficiency. To achieve effective parallelism, it is necessary to balance computation and communication while adhering to the constraints of limited accelerator memory. Identifying the optimal parallelism strategy is challenging due to varying computational and communication efficiencies across different environments, which creates a vast and complex search space. Large-scale experiments are typically required to assess these efficiencies, especially for communication operators. However, these experiments can be impractical due to the extensive search space and associated high costs. Moreover, the training setting is inherently dynamic during the overall training process, with factors like limited batch sizes in the initial stages and context-window extensions continuously altering the search space. This complexity is further intensified by the emergence of new hardware platforms, where frequent optimizations impact both efficiency and memory usage. As a result, the optimal parallelism strategy and search space are in a state of constant flux. Therefore, it is imperative to develop adaptive and continuously evolving parallelism strategies, particularly in the context of early-life hardware platforms.

\begin{figure}[!t]
  \centering
  \includegraphics[width=0.7\linewidth]{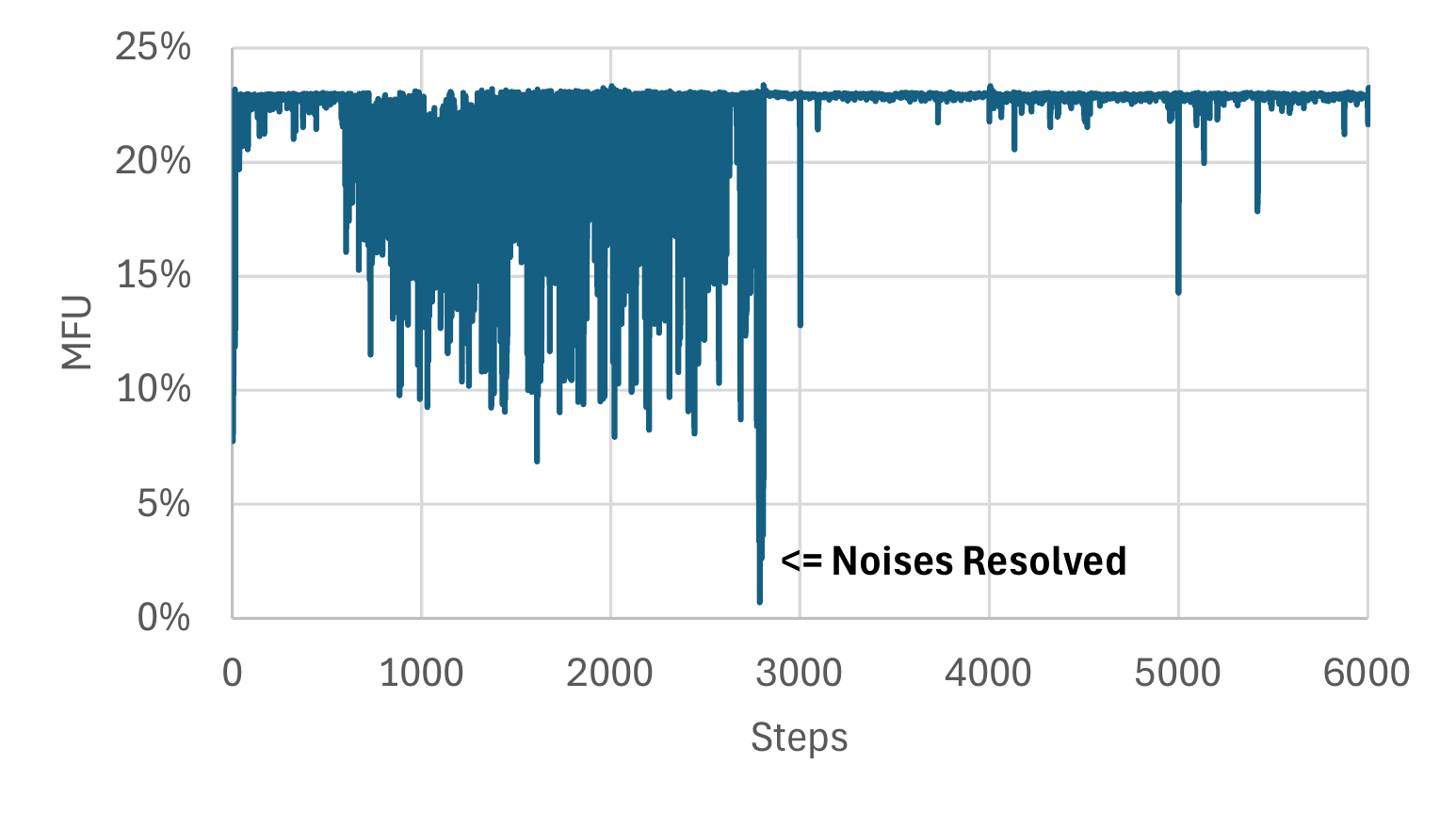}
\caption{Training MFU: Before and After Resolving Noises.}
\label{fig:efficiency-noise-motivation}
\end{figure}

\paragraph{Noises in Large-Scale Training.} The stability of training throughput is crucial, particularly for distributed training. This type of training necessitates synchronized execution of operators across AI accelerators due to algorithmic requirements. However, local disruptions such as memory management, data I/O, and scheduling can differ across AI accelerators, generating local noises and permeating the system and diminishing overall throughput. Moreover, large-scale runs exacerbate the impact of these disruptions, leading to greater fluctuations in training throughput. \cref{fig:efficiency-noise-motivation} shows an example about the significant MFU difference with and without such noises.

%% file: sections/platform.tex
\section{Lucia Training Platform}
The Lucia Training Platform (LTP) is an AI-powered cluster management system, built upon the OpenPAI project~\citep{microsoft2018openpai}, aimed at optimizing accelerator utilization and ensuring high reliability for large-scale training.

To address the challenges posed by frequent system disruptions, the LTP integrates a comprehensive reliability management system. The design's primary goal is to maximize reliability, which in turn minimizes wasted accelerator hours and builds user trust. We identified two principal sources of wasted accelerator hours:
\begin{enumerate}
    \item \textbf{Job Failure Overhead:} Time lost due to failures and subsequent recovery actions. 
    To minimize job training overhead from system disruptions, we analyzed the lifecycle of a typical large-scale training job that encounters a failure. The sequence often proceeds as follows: a job is submitted, loads its data and checkpoints, begins the training phase, and is then abruptly interrupted by a system disruption. What follows is a costly, often manual recovery process involving fault detection, faulty node isolation, and eventual job resume. This lifecycle clearly illustrates two primary sources of wasted accelerator hours: (1) the loss of productive training progress before the interruption, a cost directly related to the frequency of failures (job MTBF), and (2) the extended downtime during the reactive detection and recovery process. This led to our two-pronged strategy: Proactive Node Healthy Validation; Proactive Fault Detection and Job Recovery.
    \item \textbf{Node Downtime:} Time consumed by node diagnosis, repair, and validation after a faulty node is detected. To address the issue of minimizing the downtime of the faulty node itself to maximize fleet availability, we implemented a fully automated, closed-loop node remediation lifecycle that drastically reduces the Mean Time To Repair (MTTR).
\end{enumerate}

To tackle accelerator hour waste, we developed three key components: Proactive Node Healthy Validation for improved job MTBF through healthy-aware node scheduler allocation; Proactive and Agile Fault Detection and Job Recovery to minimize job downtime through rapid fault identification and automated recovery; and an Automated Node Remediation Lifecycle to reduce node downtime via seamless integration with Azure's internal systems for swift hardware repair and validation. Besides that, LTP also provides the chatbot-enhanced user experience, which simplifies interactions and empowers users with intuitive tools for managing complex distributed training environments effortlessly. Furthermore, LTP features a chatbot-enhanced user experience that streamlines interactions and offers intuitive tools for managing complex distributed training environments.This user-friendly design allows LTP to quickly scale and operate thousands of AI accelerators smoothly.

\input{sections/platform_cluster_management.tex}

\input{sections/platform_copilot.tex}
\input{sections/platform_results.tex}

%% file: sections/platform_cluster_management.tex
\subsection{Proactive Node Healthy Validation.} 
\label{sec:validation}

Enhancing job MTBF is critical, as we observed that with early-life AI accelerators, many training jobs fail within the first 5–20 minutes—often during data loading or checkpoint restoration, before computation even begins. This indicates that some nodes may contain pre-existing hardware issues. In large-scale synchronous training jobs, the overall job's time-to-failure is dictated by the 'weakest link,' where a single node with a shorter time-to-failure can compromise the entire job. These observations align with our previous work, \textit{SuperBench}~\citep{xiong2024superbench}, which emphasizes the importance of proactive hardware validation. To address these challenges, we developed and integrated a \textbf{proactive validation system}. Before a node is made available or allocated to a job, it undergoes either a comprehensive or a quick validation process. The job scheduler then allocates only validated nodes, ensuring improved platform availability and training reliability.

\subsection{Proactive and Agile Fault Detection and Job Recovery}

Minimizing job downtime is critical in large-scale AI systems, as system disruptions are inevitable and simply reducing failure frequency is insufficient if recovery remains slow and dependent on manual intervention. 
Job downtime is defined as the interval between a fault interrupting a job and the resumption of training on healthy resources. Any delay here directly translates to significant idle accelerator hours of all involved AI accelerators.

The first step is to identify the job is abnormal.
Following a failure, jobs often hang for extended periods—such as waiting for NCCL timeouts exceeding 30 minutes or suffering severe throughput degradation—rather than terminating promptly. Traditional platforms rely on users to manually identify and report such issues, causing significant delays in fault notification and a drop in effective resource utilization.
The second and most critical step in this process is to detect and locate the faulty machine as quickly as possible. 
The need for rapid detection is particularly challenging, a fact well-documented in existing literature on AI systems. Researchers have shown that failures are diverse, and require correlating vast amounts of telemetry data for accurate diagnosis. One new issue emerge with early-life AI accelerator platforms: undocumented system disruptions continuously arise and evolve in tandem with upgrades to the hardware and software stack.

To address these challenges, the reliability system must be agile. For known issues, agility means proactively detecting faults instead of waiting for job hang or regression and isolating affected nodes with minimal latency, followed by automated job recovery. For unknown issues, agility requires rapid discovery of new fault in the job, identification and isolation of faulty nodes, and swift conversion of unknown faults into known patterns to enable future automation.
To meet these requirements, we designed a hierarchical, three-layer architecture: Layer 1 collects and provides raw telemetry data; Layer 2 analyzes the data to discover the fault; Layer 3 automates the application of these insights for future fault detection.

\begin{itemize} 
    \item \textbf{Layer 1: The Data Foundation.} The foundation for any agile diagnosis is having the right data at the right time. Because the metrics needed for diagnosis change as system disruptions evolve, a rigid, hard-coded collection agent is too slow. Our foundational layer is therefore a \textit{declarative telemetry collection system}. It decouples the data specification from the collection logic, allowing operators to add or change collected metrics on-the-fly via a simple YAML configuration.  Using a simple YAML configuration, operators can define what data to collect from diverse sources—including driver logs, OS logs, AI accelerator, PCIe, InfiniBand counters. The system automatically collects, aggregates, and streams this data to a high-speed real-time data store, enriched with metadata like node and job IDs. This layer's sole purpose is to provide a rich, adaptable stream of data to the analysis layers above, reducing the time-to-data for new investigations from weeks to minutes.

    \item \textbf{Layer 2: The Discovery and Isolation Engine.} Building on the foundation of agile data collection, the second layer is responsible for discovering anomaly and isolating them to specific nodes. The process begins with a high-level trigger: \textit{similarity-based job anomaly detection}. Our engine performs real-time comparisons of key job-level KPIs against its historical baselines on same job with different scale. A significant deviation—for example, a drop in training throughput—flags a potential issue within the group of nodes running the job. 
    For known issues, \textit{rule-based detection} rapidly identifies faulty nodes. For unknown issues, an \textit{LLM-driven Intelligent Diagnosis Agent} automates telemetry exploration, iteratively comparing granular metrics across nodes to hypothesize and confirm the root cause. This process continues until a single anomalous node is confidently identified, providing high-precision input for automation and learning in Layer 3. Once a fault is detected and isolated, the system automatically initiates job recovery—restarting affected jobs on healthy resources and isolating faulty nodes—ensuring minimal downtime and eliminating the need for manual intervention.

    \item \textbf{Layer 3: The Automation and Learning Engine.} 
    The final layer transforms newly discovered "unknown" anomalies into automatable fault signatures for future recurrences. A dynamic \textit{Failure Signature Knowledge Base} maintains all detection rules of known issues. Traditionally, maintaining and optimizing this knowledge base is manual and slow, requiring expert validation and intelligence. To overcome this bottleneck, we employ an \textit{Agentic Time-Series Anomaly Detection framework}, leveraging Large Language Models as automated domain experts. The agent explores metrics, generates robust rules, validates them against historical data, and deploys them into the knowledge base.  This creates a feedback loop: Layer 2’s node-level isolation provides focused input for rule generation, enabling the system to learn new fault signatures and detect future recurrences instantly, without repeated exploratory diagnosis.

\end{itemize}

Although the three-layer architecture offers a strong foundation for fault management, its effectiveness in agility and scalability ultimately depends on the system’s ability to quickly respond to and learn from new failures. To address this, we introduce two AI-driven capabilities that enhance and automate the intelligence lifecycle: an online engine for real-time isolation of unknown faults and an offline engine for autonomous rule generation. The detailed design is presented in Appendix~\ref{app:fault_management}.

\subsection{Automated Node Remediation Lifecycle}

Once a node is confirmed as faulty by our detection system, the remediation process begins instantly and without human intervention. Through deep integration with the Kubernetes API, the platform automatically performs a `node cordon` operation to isolate it from the scheduler and a `job restore` operation to reschedule any affected workloads onto healthy nodes. This immediately contains the impact of the failure.

Following isolation, a policy-driven remediation engine executes a pre-defined "playbook." For software faults, this may trigger a simple reboot. However, the most significant innovation addresses uncorrectable hardware failures. Traditionally, this was a slow, multi-day process with manual handoffs: our operators were responsible for detecting the failure and filing a support ticket, and the Azure HPC/AI team was then responsible for manually verifying the issue, taking the physical host out of rotation, and performing the repair.

To eliminate this bottleneck, we co-designed a shared, automated workflow with the Azure HPC/AI team, enabled by deep API integration between our platform and Azure's internal engineering systems (ICM, Azure DevOps). In this new model, our platform takes the first step: upon diagnosing an uncorrectable hardware failure, it automatically opens a high-priority incident ticket via API, pre-populated with trusted diagnostic data. This ticket is then automatically ingested by the Azure HPC/AI team's internal systems. Trusting the provided data, their automation immediately deallocates the faulty physical host and triggers the repair or manual diagnosis process, allowing to migrate our VM to a healthy host without manual intervention and long delay.

Finally, to close the loop, our platform detects that the VM has been migrated. It then automatically subjects the "new" node to a rigorous validation suite in \ref{sec:validation} to confirm its stability and performance. Only after successfully passing this validation is the node's cordon removed, making it available for new workloads.

%% file: sections/platform_copilot.tex
\subsection{AI-based Companion for Customer Facing}

To enhance user experience and leverage the platform's robust backend capabilities, an AI-powered Companion was developed within LTP. This Companion is designed to democratize access to operational data and automate diagnostics, thereby boosting productivity for both cluster administrators and AI researchers through a natural language interface. For interactive data exploration, Companion allows administrators to convert natural language queries into precise, executable queries for various data back-ends, facilitating dynamic interaction with cluster data without needing specific query language expertise. In terms of intelligent diagnostics, Companion aids AI infrastructure specialists and researchers by identifying training instabilities, providing detailed analyses, and making the troubleshooting process transparent and informative. It gathers relevant context for diagnosing failures and translates technical errors into understandable insights, fostering collaborative problem-solving, enhancing user trust, and accelerating research progress.

During the development of Companion, one of our biggest challenges was achieving overall user satisfaction while managing the trade-off between accuracy and latency. The effectiveness of the Companion system relies on its ability to provide accurate responses within an acceptable time-frame, and user expectations vary widely based on the complexity of their questions. A one-size-fits-all approach with a single model cannot satisfy all users due to differing response time needs. Our solution involves a tiered processing strategy that integrates a complexity classifier and a performance validator to continuously improve user satisfaction. Initially, an AI-powered classifier quickly evaluates the complexity of user input to determine the optimal processing path. Simple queries are handled through efficient, low-resource methods, while more complex questions are routed to a resource-intensive multi-agent orchestration layer that delivers high accuracy, albeit with longer processing times. To maintain efficiency and high user satisfaction, the performance validator iteratively gathers data to establish benchmarks for accuracy and latency. This data-driven approach enables ongoing improvement and optimization of the system.

%% file: sections/platform_results.tex
\subsection{Cluster Operating Results}

From March 2025 to early August 2025, the platform continuously served a large-scale AI training jobs with 1,688 $\sim$ 2,144 accelerators. During this period, we progressively integrated automated fault detection, isolation, and recovery techniques, resulting in steady improvements in training reliability and operational efficiency.

\textbf{Reliability Evolution.} Initially, frequent system disruptions led to repeated job interruptions and significant accelerator resource waste. With the deployment of the proactive validation system, faulty nodes were promptly identified and isolated before job allocation, effectively renewing the node-level MTBF and reducing the risk of early failures. As a result, the average job time-to-failure for large-scale jobs ([1440, 2048] accelerators) improved substantially. The maximum single job time-to-failure increased from 24.28 hours in March to 79.87 hours in July (see Table~\ref{tab:job_ttf}), reflecting a marked reduction in job interruptions and a major advance in operational reliability.

\begin{table}[ht]
\centering
\caption{Maximum Single Job Time-to-Failure on [1440, 2048] Accelerators Per Month}
\begin{tabular}{@{}cccccc@{}}
\toprule
Month & 2025-03 & 2025-04 & 2025-05 & 2025-06 & 2025-07 \\ \midrule
\begin{tabular}[c]{@{}c@{}}Max Job Time-to-Failure (Hours)\end{tabular} & 24.28 & 37.32 & 25.97 & 56.01 & 79.87 \\ \bottomrule
\end{tabular}
\label{tab:job_ttf}
\end{table}

\textbf{Downtime Reduction.}
The automation of proactive fault detection and job recovery resulted in a dramatic reduction in job downtime. As shown in Table~\ref{tab:job_recovery}, the average time from job interruption to resumption decreased from 2.52 hours in March—when recovery relied entirely on manual intervention by cluster operators—to less than 10 minutes in July, after the deployment of rules-based automated detection and recovery starting in April. Over this period, the system continuously learned and integrated new fault signatures into its knowledge base, enabling it to automatically handle an expanding range of hardware issues. By July, most system disruptions could be detected and recovered without human intervention, greatly improving operational efficiency and user experience.

\begin{table}[ht]
\centering
\caption{Average Job Recovery Time Per Month}
\begin{tabular}{@{}cccccc@{}}
\toprule
Month & Mar-2025 & Apr-2025 & May-2025 & Jun-2025 & Jul-2025 \\ \midrule
Average Job Recovery Time (Hours) & 2.52 & 0.37 & 0.75 & 0.11 & 0.16 \\ \bottomrule
\end{tabular}
\label{tab:job_recovery}
\end{table}

\textbf{Node Remediation Efficiency.}
The deployment of automated node recycling workflows led to a dramatic acceleration in remediation speed. Node repair time is primarily bottlenecked by external AI accelerator servicing schedules. For the purpose of this analysis, we consider a theoretical best-case scenario: assuming immediate availability of replacement cluster resources. As shown in Table~\ref{tab:recycle}, after full automation was implemented in July, the average total node recycling time dropped sharply from 55.54 hours to approximately 5 hours, enabling much faster fleet recovery and substantially improving overall resource availability. Under such conditions, the entire node recycling process could be reduced to as little as 5 hours. This significant reduction in turnaround time minimizes the duration that faulty nodes remain offline, thereby maintaining higher operational capacity and mitigating the impact of hardware failures on ongoing jobs.

\begin{table}[ht]
\centering
\caption{Average Node Recycling Time (Hours) Breakdown Per Month}
\begin{tabular}{@{}ccccccc@{}}
\toprule
Month & Mar-2025 & Apr-2025 & May-2025 & Jun-2025 & Jul-2025 & Aug-2025 \\ \midrule
\begin{tabular}[c]{@{}c@{}}Disruption Detection Time\end{tabular}            & 26.79   & 15.74   & 11.66   & 6.67    & 1.09    & 0.59    \\
\begin{tabular}[c]{@{}c@{}}Node Validation Time\end{tabular}   & 28.75   & 15.34   & 17.23   & 10.04   & 4.23    & 4.45    \\
\begin{tabular}[c]{@{}c@{}}Node Recycling Time\end{tabular}     & 55.54  & 31.08  & 28.89   & 16.71   & 5.32   & 5.04   \\ \bottomrule
\end{tabular}
\label{tab:recycle}
\end{table}

\textbf{Resource Utilization.} The reliability framework drove a marked increase in overall accelerator availability, rising from 86.3\% in the first month to 98.3\% in the last month, thereby maximizing the cluster's productive capacity (see Table~\ref{tab:utilization}). 
In addition, the effective utilization of allocated accelerator hours—measured by the proportion of non-idle time—remained consistently above 99\% in the last 2 month. This sustained high utilization underscores the impact of automation in proactive fault detection and job recovery, ensuring that nearly all allocated resources contributed directly to productive computation. It is noteworthy that approximately 5\% of resources remained unutilized. This is primarily attributable to users reserving backup capacity to safeguard against sudden spikes in system disruptions, which could otherwise disrupt large-scale workloads.

\begin{table}[ht]
\centering
\caption{Cluster Resource Utilization Per Month}
\begin{tabular}{@{}cccccc@{}}
\toprule
Month                        & Mar-2025  & Apr-2025  & May-2025  & Jun-2025  & Jul-2025  \\ \midrule
Total Accelerator Hours              & 1,401,204 & 1,493,026 & 1,543,680 & 1,492,224 & 1,543,680 \\
Available Utilization \% & 86.28   & 96.13   & 97.18   & 98.99   & 97.82   \\
Allocated Utilization \% & 13.28   & 85.31   & 85.97   & 95.26  & 94.54   \\
Effective Utilization \% & 11.96 & 85.26 & 85.88 & 95.11 & 94.45 \\ \bottomrule
\end{tabular}
\label{tab:utilization}
\end{table}

\textbf{Automation Coverage and Escalation.}
The effectiveness of the reliability system is further demonstrated by its high automation coverage and minimal escalation rate. Throughout the evaluation period, the proportion of hardware failure incidents resolved automatically—without human intervention—increased sharply, The monthly automation recovery ratio rose from 43.47\% in March to 97.83\% in July as shown in Table~\ref{tab:automation_ratio}. 
This sustained automation not only ensured rapid recovery and uninterrupted large-scale training, but also significantly reduced the operational burden on human operators, allowing less operating effort to manage the cluster while maintaining high reliability.

\begin{table}[ht]
\caption{Monthly Job Hardware Failure Automation Recovery Ratio}
\centering
\begin{tabular}{@{}cccccc@{}}
\toprule
Month                  & Mar-2025 & Apr-2025 & May-2025 & June-2025  & Jul-2025  \\
Automation Ratio (\%) & 43.47    & 87.50    & 91.67    & 96.00      &  97.83    \\ \bottomrule
\end{tabular}
\label{tab:automation_ratio}
\end{table}

\textbf{Companion Accuracy and Response Time.}
We evaluated the accuracy and response time of our deployed Companion using a validation dataset consisting of 90 questions. As shown in Table \ref{tab:companion_performance_latency}, the accuracy increased significantly from 45.6\% to 87.8\%, an improvement by a factor of 1.93. Meanwhile, the average response time for simple questions experienced a slight increase due to classifier overhead. This considerable improvement in overall accuracy, along with a consistent experience for simple questions, greatly enhances the overall user experience.

\begin{table}[ht]  
    \centering  
    \caption{Companion Accuracy and Response Time}  
    \label{tab:companion_performance_latency}  
    \begin{tabular}{@{\hskip 0.1in}llcc@{\hskip 0.1in}}  
        \toprule  
        \multicolumn{2}{l}{\textbf{Metric}} & \textbf{Baseline} & \textbf{LTP Companion} \\ \midrule  
        \multirow{3}{*}{Average Response Time (s)}   
        & Simple & 4.42 & 5.82 \\  
        & Hard & 4.29 & 11.08 \\
        & Overall & 4.34 & 9.15 \\  
        \midrule  
        \multicolumn{2}{l}{Accuracy} & 45.6\% & 87.8\% \\   
        \bottomrule  
    \end{tabular}  
\end{table} 

%% file: sections/framework.tex
\section{Lucia Training Framework}

The Lucia Training Framework (LTF), built upon Megatron-LM~\citep{shoeybi2019megatron}, is a unified and extendable training framework designed to enable stable and efficient training on early-life hardware platforms.

\input{sections/framework_stability}
\input{sections/framework_efficiency}
\input{sections/framework_results.tex}

%% file: sections/framework_stability.tex
\subsection{Stability Improvement}\label{sec:stability}

In this section, we present two key technologies to tackle the stability challenge mentioned in \cref{subsec:challenge-stability}.

\begin{table}[!t]
\centering
\caption{Module-Level Cosine Similarity Between Different Software Stacks}
\begin{tabular}{@{}cccccc@{}}
\toprule
Modules & Outputs & Parameters & Gradients  \\ \midrule
Attention & 0.9998 & \textbf{0.1626} & \textbf{0.5815} \\
Bias-Dropout-Add & \textbf{0.9309} & - & \textbf{0.9000} \\
Embedding & \textbf{0.8995} & 0.9993 & \textbf{0.9142} \\
MoE (EP=8) & 0.9998 & 0.9994 & \textbf{0.9484} \\ \bottomrule
\end{tabular}
\label{tab:module-similarity}
\end{table}

\paragraph{Proactive Numerical Validation} Proactive numerical validation before each training session is crucial for maintaining stability. By conducting these validations, researchers can concentrate on addressing model design flaws when stability issues arise, thereby significantly reducing diagnostic time. We developed a numerical test pipeline to validate all components used in the training workload. Specifically, based on the requirements of AI researchers, components on new hardware platforms are expected to produce outputs and gradients that closely match those on established platforms when evaluated using both random and real input samples and weights. We also compare further optimizer steps to eliminate accumulated errors. Numerical passing thresholds for different operators are determined heuristically based on user requirements. \cref{tab:module-similarity} lists the modules identified during the proactive numerical validation of \modelname{} training on different software stacks with different accelerators. We assessed the minimum cosine similarities between outputs, parameters, and gradients after 10 optimizer steps on both platforms, marking any result below 0.99 as abnormal. Further analysis revealed that the Attention module error resulted from a one-time error in the qk-layernorm layer and a 10-step accumulated error in the output linear layer's bias. The MoE layer error was due to a 10-step accumulated error in the MoE router. Additionally, the errors in the bias-drop-add and embedding layers were caused by differing random number generator implementations across the hardware. Proactive numerical validation allows us to address correctness issues in advance, ensuring the end-to-end training stability.

\begin{figure*}[t]
    \centering
    \hspace{-0.015\textwidth}
    \includegraphics[width=0.99\textwidth]{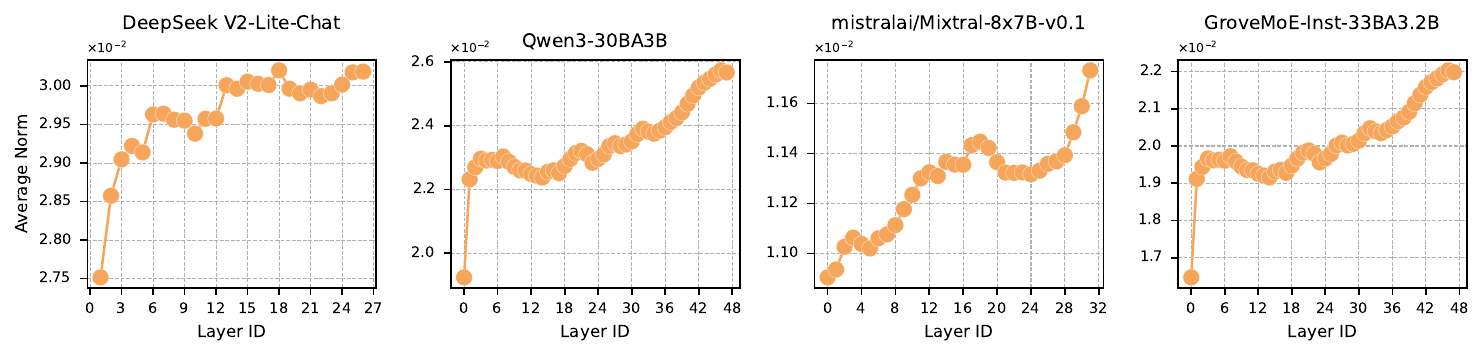}
    \caption{MoE parameter average norm of different open-sourced models.}
    \label{fig:ffn_scale}
\end{figure*}

\paragraph{Early and Accurate Training Collapse Detection} To better predict the potential model collapse, we compare successfully trained models with crashed models, reveal significant differences in their parameter norm distributions, and identify a key condition potentially essential for stable training: the averaged norm of the MoE parameters should generally increase with deeper layers. We begin by comparing the parameter norm distributions of a model that experienced training collapse with those of several open-source models, including DeepSeekV2-lite, Qwen3-30BA3B, Mixtral-MoE, and Grove-MoE. Specifically, for each model, we compute the norm of the MoE parameters across layers, and visualize the layer-wise parameter norm in~\cref{fig:ffn_scale} and~\cref{fig:crashed_moe_scale}. As shown in the figures, all open-source models exhibit a pattern where the parameter norm generally increases with layer depth, whereas the collapsed model displays a distinct valley phenomenon, with the parameter norm of the middle layers significantly smaller than those of both shallow and deep layers. These observations suggest that an increasing parameter norm across layers may be a necessary condition for successful pre-training. Based on this observation, we can detect training collapse accurately and in advance.

%% file: sections/framework_efficiency.tex
\subsection{Efficiency Improvement}
In this section, we introduce two key technologies designed to address the efficiency challenge outlined in \cref{subsec:challenge-efficiency}.

\subsubsection{AI-Assisted Noise Detection.}
We utilize the extensive domain knowledge embedded in state-of-the-art large language models (LLMs) to aid in detecting the straggler phenomenon during synchronization, which is caused by local disruptions. With the assistance of AI, we have identified two cases, previously unknown to us, through insights provided by LLMs.
\paragraph{Lazy OS Page Cache Invalidation.} To load data efficiently, Megatron-LM's data loader allows the operating system (OS) to keep indices of underlying raw datasets in the page cache, consuming a majority of available host memory when blending many datasets. This becomes problematic when a job is rescheduled and ranks are remapped to different nodes. The job then starts with host memory occupied by page cache from previous runs, and the OS interrupts job processes to recycle obsolete page cache in a lazy and random manner, causing persistent local disruptions. To mitigate this, we manually drop all cache before each job run.  
\paragraph{Python Garbage Collection (GC).} Model training involves frequent memory allocation activities on both CPU and accelerator. In the Python runtime, garbage collection (GC) is triggered frequently but inconsistently across different ranks, leading to random overhead at varying times. We leverage Megatron-LM's option to disable Python's automatic GC and perform manual GC at fixed intervals.

\subsubsection{Pruned Parallelism Tuning}
Parallelism tuning is crucial for improving overall peak throughput, but the search space is expansive, making it impractical to perform on a large scale. Therefore, we begin by performing a comprehensive search and analysis across the four dimensions of parallelism—Expert Parallelism (EP), Tensor Parallelism (TP), Pipeline Parallelism (PP), and micro batch size (MBS) on a small scale, utilizing 16 AI accelerators. \footnote{Since the global batch size is determined by AI researchers based on training algorithm considerations, Data Parallelism (DP) is automatically determined once the number of accelerators, global batch size, and other parallelism dimensions are established.} This helps identify constraints related to hardware topology, software stack efficiency, and model design, which can reduce the tuning search space. After establishing these constraints, we can automatically search the refined space in real environments and settings. We provide some examples of training combinations to illustrate our comprehensive tuning method on different platforms.

\begin{itemize}  
    \item \textbf{Heavy Communication Overhead on EP and TP.} From a software–hardware co-optimization perspective, a full-mesh topology offers strong global connectivity and high aggregate bandwidth, which are advantageous for large-scale collective operations. However, even within a full-mesh design, the effective bandwidth available to different subgroup sizes may differ from the peak full-node bandwidth. This does not diminish the benefits of full-mesh connectivity; rather, it underscores the need for software to adopt sharding and communication strategies that match the bandwidth characteristics of each subgroup to fully capitalize on the hardware. Without such alignment, EP and TP performance can be constrained by communication patterns that are not topology-aware. For \modelname{}, this consideration directly informs parallelism choices. The exploration of TP and EP configurations shows that TP=1 and EP=8 best leverage the corresponding topologies and communication behaviors in the current software stack.
    \item \textbf{Tradeoff between PP and MBS.} Virtual Pipeline Parallelism (VPP) allows each PP stage to be divided into substages, enabling an increase in PP to reduce memory overhead on each accelerator with limited performance regression. Simultaneously, the increased memory capacity allows for a larger MBS, which, according to our experiments, can enhance end-to-end performance by up to 30\%. Therefore, the tradeoff between PP and MBS should be explored in large-scale real-world experiments.
\end{itemize}

In conclusion, we performed comprehensive small-scale testing on each training combination for new AI accelerators to narrow down the research space for parallelism tuning. This approach enables us to identify the optimal configuration in a real large-scale environment by considering only limited MBS and model parallelism options, while adhering to memory constraints. This pruned parallelism tuning allows us to determine the best parallelism setup in a very short time.

%% file: sections/framework_results.tex
\subsection{Results}
From May 2025 to early August 2025, we utilized the LTF for \modelname{} training with 2,048 accelerators. In this section, we present the improvements in stability and efficiency achieved by implementing LTF in \modelname{} training.

\begin{figure}[!t]
  \centering
  \includegraphics[width=0.8\linewidth, height=7cm]{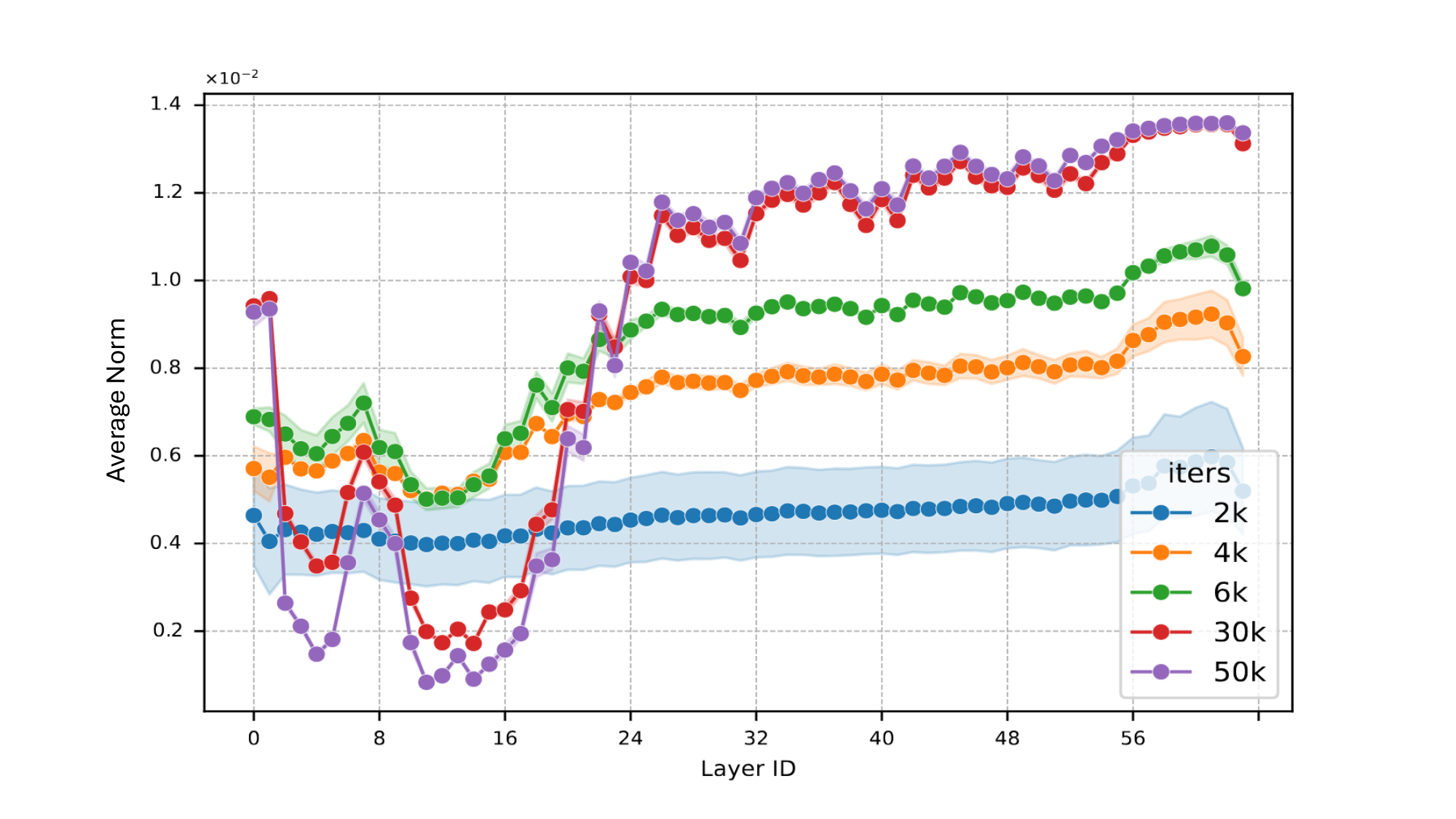}
\caption{The change in MoE parameter average norm of a crashed model over training iterations.}
\label{fig:crashed_moe_scale}
\end{figure}

\paragraph{Training Stability.} Following proactive numerical validation, the 75-day \modelname{} training encountered stability issues only \textit{once}. Thanks to parameter norm distribution based collapse detection, we identified the issue at 6K steps, which is five times earlier than the actual collapse point at 30K steps. As shown in Figure~\ref{fig:crashed_moe_scale}, we observed that once the model experienced a valley phenomenon in its MoE parameter average norm distribution early in pre-training, this issue progressively worsened as training continued, ultimately leading to model collapse.

\begin{table}[!t]  
\centering  
\caption{Efficiency Optimization Trajectory for \modelname{} Training}  
\begin{tabular}{@{\hskip 0.1in}l@{\hskip 0.2in}l@{}}  
\toprule  
\textbf{Optimizations} & \textbf{MFU} \\  
\midrule  
Initial baseline & \multirow{2}{*}{9.86\%} \\  
Parallelism: TP1-CP1-EP8-PP8, 8 layers per PP stage for a 64-layer model \\
\midrule
Fused-attention with vendor-specific backend & 10.34\% \\
\midrule
Global batch size changed from 16M to 32M & 13.04\% \\  
\midrule
Global batch size changed from 32M to 37M & 13.57\% \\  
\midrule
Fused Bias-SwiGLU & \multirow{5}{*}{13.82\%} \\   
Fused gradient accumulation \\
Fused grouped MLP GEMM with vendor-specific backend \\  
Fused RoPE \\  
Vendor-specific runtime tuning for D2H and H2D memory copy \\  
\midrule
More balanced PP: change the first stage to 9 layers and the last stage to 7 layers & 16.04\% \\ 
\midrule
More balanced PP: change model layers to 63 and the first stage to 8 layers & \multirow{2}{*}{18.78\%} \\
Fused MoE token permute/unpermute \\
\midrule
Enable VPP=2 to further split each PP stage into 2 sub-stages & \multirow{2}{*}{20.41\%} \\   
Selective MoE layer re-compute to preserve enough memory for VPP \\
\midrule
Prioritize MoE layers with more balanced routing for re-compute & 21.08\% \\  
\bottomrule  
\end{tabular}  
\label{tab:efficiency-optimizations}  
\end{table}

\begin{table}[!t]
\centering
\caption{\modelname{} training throughput before and after resolving two stragglers}
\begin{tabular}{cccc}
\toprule
\textbf{Stage} & \textbf{Min / Peak} & \textbf{Average / Peak} & \textbf{Std} \\
\midrule
Before & 29.47\% & 85.46\% & 22.80 \\
After  & 55.05\% & 98.11\% & 1.98 \\
\bottomrule
\end{tabular}
\label{tab:efficiency-noise-results}
\end{table}

\paragraph{Training Efficiency.} We embarked on a 12-week journey to enhance the end-to-end performance of the \modelname{}. As shown in \cref{tab:efficiency-optimizations}, we implemented various optimizations, including component optimization and parallelism tuning. This effort increased the end-to-end peak MFU from 10.43\% to 21.08\%. Regarding the stability of training throughput, after the two stragglers identified by the AI-Assisted Noise Detection module are resolved, the effective throughput of \modelname{} increases from 85.46\% to 98.11\% of peak performance, with a 11.5x reduction in standard deviation, as shown in \cref{tab:efficiency-noise-results}.

%% file: sections/model.tex
\section{\project{} Model Training Validation}

Using \project{} platform, we pre-train a decoder-only model named as \modelname{}, which contains 63 blocks and each block has a group query attention (GQA)~\citep{ainslie2023gqa} layer followed by a Mixture-of-Experts (MoE) layer~\citep{dai2024deepseekmoe}. More specifically, each MoE layer contains 96 experts with 8 activated per token, yielding a 200-billion-parameter model with 20 activated billion parameters (200BA20B) for each token.

\subsection{Training Stability}

\begin{figure}[!t]
    \centering
    \subfigure[Training Loss]{
    \includegraphics[width=0.475\columnwidth]{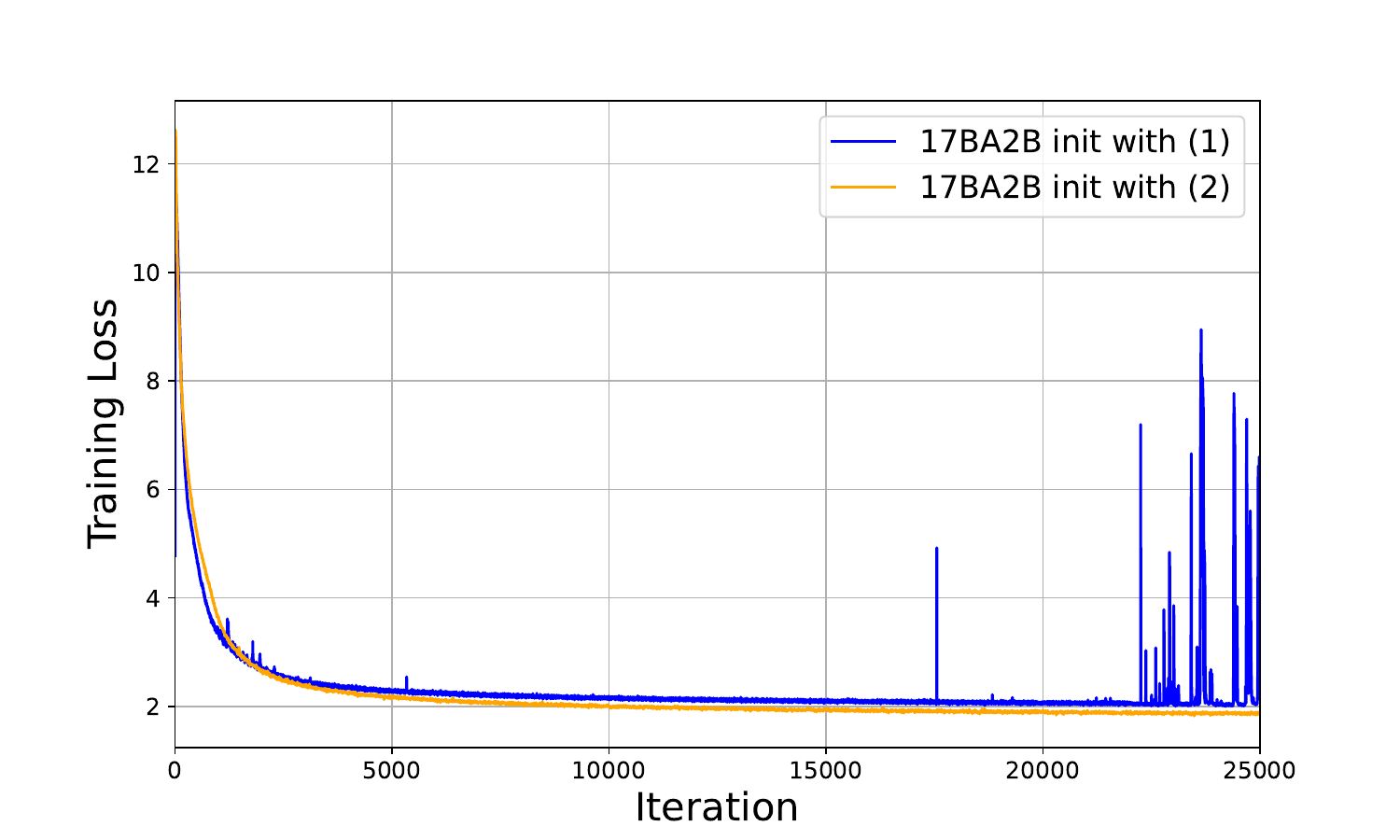}
    }
    \subfigure[Gradient Norm]{
    \includegraphics[width=0.475\columnwidth]{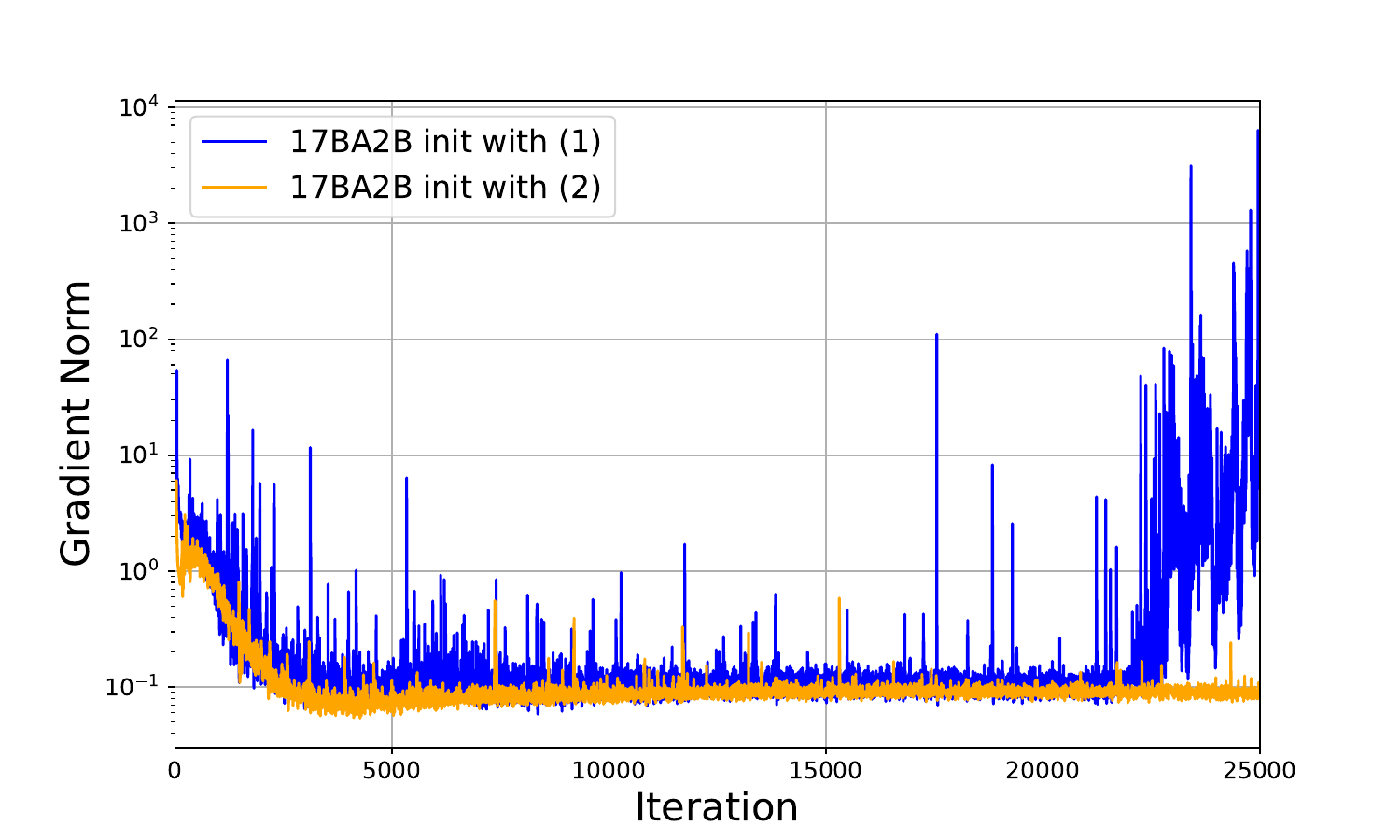}
    }
    \caption{Training loss and gradient norms across models under different initialization settings.}\label{fig:loss_grad}
    \vspace{-1em}
\end{figure}

Using the training collapse detection approach mentioned in Section~\ref{sec:stability}, we compare different choices of initialization, seeking suitable settings to prevent model collapse. Let $\sigma$ represent the user-defined initial standard deviation (init std), and $L$ denote the number of layers. We specifically compared two initialization methods:  
  
\begin{itemize}  
    \item \textbf{Scaled Output Layer Initialization}: This is the default setting used in the open-source Megatron. Here, the init std for all output layer parameters is set to $\sigma/\sqrt{2L}$, while the init std for other parameters is set to $\sigma$.  
    \item \textbf{Gaussian Initialization}: In this method, the init std for all parameters is uniformly set to $\sigma$. This approach is also utilized in DeepSeek-V2 and DeepSeek-V3.  
\end{itemize}  

We pre-trained two models using the 17BA2B architecture and analyzed their parameter norm distributions for comparison. We discovered that the model initialized using Gaussian Initialization exhibited the same parameter norm distribution as the open-source models within 5,000 iterations, whereas the model using Scaled Output Layer Initialization did not. Additionally, as shown in Figure~\ref{fig:loss_grad}, the model initialized with Gaussian Initialization demonstrated smoother training loss curves and fewer gradient norm spikes, indicating that this uniform initialization method is more suitable for our scenarios.

%% file: sections/results.tex
\subsection{Training Playground}

\input{tables/pretraining_results}
\subsubsection{Settings}
To assess the stability of our training framework (\ltf{}) on the LTP platform, we conducted the general stage pre-training\footnote{The reasoning stage and annealing procedures are not included.} and evaluated the \modelname{}-Base across a diverse set of tasks encompassing general knowledge, mathematics, and coding. The benchmarks are categorized as follows:

\begin{itemize}
  \item \textbf{General Tasks}: To assess world knowledge, we employ MMLU~\citep{mmlu}, MMLU-Pro~\citep{mmlupro}, and SuperGPQA~\citep{supergpqa}. To evaluate English reading comprehension and contextual reasoning, we adopt BigBenchHard (BBH)~\citep{bbh}, PIQA~\citep{piqa}, ARC~\citep{arc}, HellaSwag~\citep{hellaswag}, and WinoGrande~\citep{sakaguchi2021winogrande}. For assessing scientific knowledge, we use GPQA~\citep{rein2024gpqa}, which focuses on graduate-level scientific questions.

  \item \textbf{Reasoning Tasks}: We evaluate mathematical reasoning capabilities with GSM8K~\citep{gsm8k} for foundational arithmetic and MATH~\citep{math} for advanced problem solving. Code reasoning ability is assessed on HumanEval~\citep{humaneval}.
  
\end{itemize}

We compare \modelname{}-Base with representative base models, including DeepSeek-V2-Base~\citep{liu2024deepseekv2}, Qwen2.5-72B-Base~\citep{yang2024qwen25} and LLaMA-3.1-405B-Base.

\subsubsection{Results}

The evaluation results of \modelname{}-Base are presented in \cref{tab:sigma-pretrain}.

\modelname{}-Base achieves over 80\% accuracy on the MMLU benchmark~\citep{mmlu}, while demonstrating solid performance on mathematical reasoning tasks, with 54.6\% on MATH~\citep{math} and 84.1\% on GSM8K~\citep{gsm8k}.
The model also performs well on coding reasoning tasks, attaining 57.9\% pass@1 on HumanEval~\citep{humaneval}.

The above results show that our model achieves performance comparable to other baselines after pre-training, demonstrating that our training platform and framework are capable of supporting large-scale distributed training.

%% file: tables/pretraining_results.tex
\begin{table}[!t]
\centering
\footnotesize
\setlength{\tabcolsep}{4.5pt}
\begin{tabular}{@{}l l l | c c c | c@{}}
\toprule
& \multirow{2}{*}{\centering \textbf{Benchmark {\tiny (Metric)}}} & \multirow{2}{*}{\textbf{\# Shots}} & \textbf{DeepSeek V2} & \textbf{Qwen2.5} & \textbf{LLaMA-3.1} & \textbf{\modelname{}} \\
& & & \textbf{Base} & \textbf{72B Base} & \textbf{405B Base} & \textbf{Base} \\
\midrule
& Architecture & - & MoE & Dense & Dense & MoE\\
& \# Activated Params & - & 21B & 72B & 405B  & 20B\\
& \# Total Params & - & 236B & 72B & 405B & 200B\\
& \# Trained Tokens & - & 8.1T & 18T & 15T & 9T\\
\midrule
\multirow{7}{*}{\textit{General}} & BBH {\tiny (EM)} & 3-shot & 78.8 & 79.8 & \textbf{82.9} & 82.3\\
& MMLU  & 5-shot & 78.4 & \textbf{85.0} & 84.4 & 80.5\\
& MMLU-Pro  & 5-shot & 51.4 & 58.3 & 52.8 & \textbf{59.8}\\ 
& ARC-Easy  & 25-shot & 97.6 & \textbf{98.4} & \textbf{98.4} & 97.2 \\ 
& ARC-Challenge  & 25-shot & 92.2 & 94.5 & \textbf{95.3} & 91.5 \\ 
& HellaSwag  & 10-shot & 87.1 & 84.8 & \textbf{89.2} & 87.1 \\
& WinoGrande  & 5-shot & \textbf{86.3} & 82.3 & 85.2 & 82.7 \\
& PIQA  & 0-shot & 83.9 & 82.6 & \textbf{85.9} & 84.6 \\
\midrule
\multirow{3}{*}{\textit{Reasoning}} & GSM8K  & 8-shot & 81.6 & \textbf{88.3} & 83.5 & 84.1\\
& MATH  & 4-shot & 43.4 & 54.4 & 49.0 & \textbf{54.6}\\
& HumanEval {\tiny (Pass@1)} & 0-shot & 43.3 & 53.0 & 54.9 & \textbf{57.9}\\
\bottomrule
\end{tabular}
\caption{Performance of \modelname{}-Base across general knowledge and reasoning benchmarks. It achieves $>$80\% on MMLU, 54.6\% on MATH, 84.1\% on GSM8K, and 57.9\% pass@1 on HumanEval. }
\label{tab:sigma-pretrain}
\end{table}

%% file: sections/conclusion.tex
\section{Conclusion}

Large-scale training on early-life AI hardware presents significant challenges in reliability, stability, and efficiency. We introduced \project{}, a streamlined platform and training stack framework, including \ltp{} (LTP) and \ltf{} (LTF), which employs learning-driven AI techniques to address system disruptions, detect and correct instabilities, and continuously enhance training efficiency. On early-life AI hardware platforms, LTP has been operational for five months, achieving 94.45\% effective accelerator utilization. Meanwhile, LTF successfully trained \modelname{}, reaching 21.08\% MFU, state-of-the-art downstream accuracy, and experiencing only a single stability incident over a 75-day period. \project{} is released as open-source to facilitate the co-evolution of AI models and AI clusters, offering a viable alternative to prevailing established accelerator stacks.

%% file: sections/appendix.tex
\clearpage
\section*{Appendix}
\input{sections/appedix_contributions}
\clearpage
\input{sections/appendix_fault_management}

%% file: sections/appedix_contributions.tex
\section{Contributions and Acknowledgments}
\begin{multicols}{2} 
\setstretch{1.4}
\textbf{Research \& Engineering} \\
Lei Qu \\  
Lianhai Ren* \\  
Peng Cheng \\     
Rui Gao \\  
Ruizhe Wang* \\    
Tianyu Chen \\  
Xiao Liu \\
Xingjian Zhang* \\
Yeyun Gong \\
Yifan Xiong \\
Yucheng Ding* \\ 
Yuting Jiang \\
Zhenghao Lin \\
Zhongxin Guo \\
Ziyue Yang \\   
\\
\textbf{Acknowledgments for Technical Exploration} \\
Haoran Deng* \\
Jian Jiao \\
Qingguo Hu* \\
Runxi Cheng* \\  
Shuai Lu* \\
Xiao Liang* \\
Yaoxiang Wang* \\
Yelong Shen \\
Yu Yan* \\
\\
\textbf{Acknowledgments for Data} \\
Han Zhang* \\
Hao Ni* \\
Hongyi He* \\
Kailai Yang* \\ 
Mingni Tang* \\
Qinglin Zhu* \\
Qinzheng Sun* \\
Shimao Zhang* \\
Yi Cheng* \\
Yasen Hu* \\  
Ying Xin \\
Zijie Chen* \\
\\
\\
\\
\\
\textbf{Acknowledgments for Microsoft Azure} \\
Anna Daly \\
Boris Pinzur \\  
Guoshuai Zhao \\  
Hossein Pourreza \\  
Joe Chau \\  
Julia Katilevsky \\  
Logan Cope \\  
Luis Martinez Castillo \\  
Ray Jui-Hao Chiang \\  
Ryn Vinogradova
\end{multicols}

Within each category, authors are listed alphabetically by first name. An asterisk (*) next to a name indicates individuals who are interns, vendors, or those who have left the team. We offer our heartfelt thanks to Microsoft Azure for their exceptional support and collaboration on this work. Additionally, we are grateful to all those who contributed to the SIGMA project but are not mentioned in this paper.

%% file: sections/appendix_fault_management.tex
\section{LLM-POWERED AGILE FAULT MANAGEMENT}
\label{app:fault_management}

While the three-layer architecture provides a robust framework for fault management, its agility and scalability are still ultimately determined by the speed at which the system can handle and learn from new failures. 
The manual processes of isolating a fault during an incident and then engineering a new detection rule afterward are significant bottlenecks. 
To solve this, we introduce two distinct AI-powered capabilities designed to accelerate and automate the intelligence lifecycle: an online engine for real-time unknown fault isolation and an offline engine for autonomous rule generation.

\subsection{Opportunities and Challenges of LLMs for Reliability Systems}

Large Language Models (LLMs) have demonstrated exceptional capabilities in understanding complex patterns, generating human-like text, and performing advanced reasoning across diverse domains. Their potential to transform reliability systems is evident in several key aspects:

\textbf{Scalable Automation of Expert Knowledge on Large-Scale Telemetry Data:} LLMs can encapsulate extensive domain expertise, enabling the automation of complex diagnostic tasks that previously required significant human effort. This scalability allows for rapid adaptation to new failure modes and evolving system architectures without extensive manual reprogramming. The challenge is particularly acute in large-scale training environments, where a single 2,000-accelerator job can generate telemetry from thousands of accelerators—each reporting hundreds of metrics and logs every second. Manually diagnosing issues, generating detection rules, and verifying results at this scale is prohibitively labor-intensive, making LLM-driven automation especially valuable.

\textbf{LLM capable of data, log correlation analysis and Pattern Recognition:} LLMs excel at identifying patterns and correlations within large-scale telemetry datasets, leveraging highly accurate code generation capabilities. This is particularly valuable for reliability systems, where uncovering relationships among diverse telemetry signals is critical for precise fault diagnosis. Besides, LLMs can naturallly understand the log data, which traditional rule-based systems often struggle with due to the unstructured nature and only limited predefined prior knowledge by existing rules.

\textbf{Explainable Reasoning for Decision-Making:} The reasoning abilities of LLMs facilitate the generation of explainable diagnostic outputs and detection rules, enhancing transparency and trust in automated production systems. This is especially important in reliability contexts, where understanding the rationale behind decisions is essential for validation and compliance.

However, integrating LLMs into production-grade reliability systems introduces several unique challenges that require systematic solutions:

\textbf{Unpredictable Accuracy:} The accuracy of LLMs on specialized tasks can be inconsistent and may fall short of existing solutions without careful design, particularly in complex domains such as system telemetry. Achieving consistently reliable accuracy remains a significant hurdle for production deployment.

\textbf{Cost and Efficiency:} LLMs are computationally intensive, presenting an additional challenge in applying them cost-effectively at scale.

\textbf{Deployment Risks from Non-Determinism:} The inherent non-determinism of LLMs poses risks for production environments, as unvalidated outputs—such as auto-generated detection rules—could destabilize system operations if deployed without rigorous validation.

\subsection{LLM-powered Agile Fault Management}
\paragraph{LLM-driven Online Fault Isolation}
To achieve rapid fault isolation in a live production environment, we designed a collaborative multi-agent system that executes a structured, iterative diagnostic workflow. 
The critical challenge is to explore the vast telemetry space to isolate the specific source of the anomaly
It autonomously explores more granular metrics from Layer 1, comparing signals across the nodes in the anomalous group. It propose hypothesize that a specific node is the cause and then query for detailed component metrics (e.g., accelerator utilization, network counters) to confirm or deny this hypothesis.
This system navigates the high-dimensional telemetry space by orchestrating the distinct capabilities of three specialized agents, systematically addressing the core challenges of employing LLMs for real-time diagnostics.

The diagnostic process is a cognitive loop driven by the interplay of these agents:
\begin{enumerate}
    \item The \textbf{Exploration Agent} initiates the diagnostic cycle. It leverages a broad knowledge base of system failure modes to generate a high-level hypothesis regarding the fault's origin, such as proposing an investigation into network performance.

    \item The \textbf{Data Analysis Agent} is responsible for evidence gathering and serves as our primary solution to the \textit{efficiency} challenge. It translates the abstract hypothesis into executable Python code . It's guided by two simple observations in LLM training workloads: 1) \textit{Spatial Coherence}: due to the synchronous nature of training, healthy nodes exhibit statistically similar metric profiles, making deviation from the peer-group a strong anomaly signal. 2) \textit{Temporal Consistency}: a node's metrics should remain consistent with its own historical baseline from healthy runs on the same workloads. The agent thus generates code to compute anomaly scores along both a spatial and a temporal dimension, reducing a massive, noisy dataset into a concise, multi-dimensional signal.

    \item The \textbf{Reasoning and Reflection Agent} receives this structured data and addresses the \textit{accuracy} challenge. It analyzes the multi-dimensional scores to evaluate the current hypothesis. If the evidence is conclusive and reproducable—for instance, a single node presenting as a dominant outlier for multiple times—the agent forms a terminal diagnosis. If the evidence is ambiguous, it reflects on the findings and provides structured feedback to the Exploration Agent. This self-correcting loop, where each iteration is informed by the last, allows the system to methodically converge on the root cause rather than relying on a single, high-risk inference.
\end{enumerate}

Finally, the workflow concludes with a critical verification stage to solve the \textit{correctness and deployment risk}. The agent system's final diagnosis is treated as a high-confidence, but still unverified, hypothesis. It is validated empirically: the identified node is programmatically isolated, and the system monitors for a corresponding restoration of job performance. This real-world feedback provides definitive confirmation of the diagnosis's correctness.

\paragraph{Autonomous Offline Time-Series Anomaly Rule Generation}
After a new type of failure occurs, we need to teach our system to recognize it instantly in the future. To do this automatically, we built a \textbf{Training Engine} that creates new detection rules. This workflow is explicitly designed to transform the unpredictable nature of LLMs into a reliable, convergent process.

The workflow begins with a crucial pre-processing stage designed to solve the challenge of \textit{efficiency}. To avoid the high cost of operating LLMs on large, noisy datasets, we first employ strategic data and feature selection. We perform contrastive feature selection by calculating feature importance between the anomalous data and a normal baseline to identify the most salient metrics. Concurrently, contextual data selection retrieves historical time-series data from jobs that are behaviorally similar to the incident but have a contrasting ground-truth label. These steps provide the LLM with a concise, high-signal, and contextually relevant dataset, maximizing its reasoning capability while minimizing computational overhead.

Next, the curated data enters a multi-agent feedback loop designed to solve the challenges of \textit{accuracy} and \textit{correctness}. This loop is inspired by backpropagation and involves three agents:
\begin{itemize}
    \item The \textbf{Detection Agent} initiates the loop by proposing a set of detection rules based on the data, formatted as Python code, based on its analysis of the input data.
    \item The \textbf{Repair Agent} then ensures \textit{syntactic correctness}.  It validates the proposed rules by executing them against a pre-defined data schema. If errors are found, uses compiler feedback to debug and fix them. This step guarantees the rule is executable, de-risking the first layer of deployment uncertainty.
    \item The \textbf{Review Agent} subsequently ensures \textit{logical correctness} and improves \textit{accuracy}.  It evaluates the syntactically valid rules against a hold-out validation dataset. A key constraint of this stage is preventing accuracy regression; the performance of any new rule set must be greater than or equal to the previous iteration's. If accuracy degrades, the agent is provided with a diagnostic report containing code differences, performance metrics, and specific data samples that were misclassified. This detailed feedback allows the system to iteratively refine the rule's logic. This transforms the model's unpredictable accuracy into a measurable, convergent process.
\end{itemize}
This structured, multi-agent feedback loop mitigates the inherent uncertainty of LLMs by transforming rule generation into a deterministic workflow with verifiable pipeline.
Only rules that pass this multi-layered validation are committed to the Failure Signature Knowledge Base.  
This automated system reduces a multi-day manual engineering task to a workflow that completes in minutes, enabling the platform to adapt to novel hardware failure modes with unprecedented speed and reliability.

\subsection{Case Study}

The AI-powered fault management system is currently under active optimization and has not yet been deployed online; instead, it is used offline to assist OFR engineers and reduce manual effort. Below, we present 2 simple representative cases where the offline system successfully supported fault diagnosis and rule generation.

\textbf{Case 1: Autonomous Anomaly Rule Generation of Accelerator Memory Fault}

The memory access fault was first observed in April 2025 during a large-scale training job on 2,048 accelerators. The job experienced intermittent accelerator hangs, with errors such as "Memory access fault by Node-9 (Agent handle: 0xxxxxxxx) on address 0xxxxxxxx. Reason: Unknown." logged on the faulty node. This error can be triggered either by user-induced temporary out-of-bounds memory access which will not affect following jobs or by hardware and driver disruptions, including DRAM ECC errors, unhealthy High-Bandwidth Memory (HBM), driver hang, etc. which may hang the node and all following jobs, incorrectly calculate loss into NaN without node cold reboot or even return to repair.
To address this, we applied the Autonomous Offline Time-Series Anomaly Rule Generation workflow. Feature selection identified job logs and dmesg logs as the most informative signals distinguishing true system disruptions from normal samples. During data selection, priority was given to collecting user-triggered memory access faults with similar log messages, ensuring both user and hardware-triggered cases were included for rule training. After three iterations with the correction of repair and review agent, the Rule Generation Agent produced detection rules that accurately identified hardware-induced faults without misclassifying user-triggered errors within one hour. The final rule combined log pattern matching with node ranking based on time-series DRAM metrics. This case demonstrates the system's ability to autonomously generate precise detection rules for complex fault scenarios, significantly reducing manual engineering effort and expertise required.

\textbf{Case 2: LLM-driven Fault Isolation for Accelerator Loss NaN Issue}

Another representative case is the identification of a Loss NaN issue during a large-scale training job in July 2025. The job, running on 2,048 accelerators, experienced abrupt termination, with many ranks reporting "found NaN in local forward loss calculation" errors. This novel error propagated rapidly across the cluster, making it challenging to pinpoint the source node.
Leveraging the LLM-driven Online Fault Isolation workflow, the system initiated a diagnostic cycle. The Exploration Agent hypothesized that the root cause could be related to accelerator compute or memory faults. The Data Analysis Agent generated Python code to compute spatial and temporal anomaly scores across accelerator core metrics (e.g., compute utilization), memory metrics (e.g., HBM error counts, ECC error counts, memory bandwidth). The Reasoning and Reflection Agent evaluated the results, initially identifying seven nodes with elevated anomaly scores with early dropped compute utilization. Through further iterations, the system incorporated historical metrics and logs during recent jobs, ultimately converging on a single node as the most likely culprit, based on its repeated abnormal log patterns and earliest time correlation with the NaN events.
This diagnosis was validated by isolating the identified node, which confirmed the effectiveness of the approach.

%% file: iclr2026_conference.bib
@article{ainslie2023gqa,
  title={Gqa: Training generalized multi-query transformer models from multi-head checkpoints},
  author={Ainslie, Joshua and Lee-Thorp, James and De Jong, Michiel and Zemlyanskiy, Yury and Lebr{\'o}n, Federico and Sanghai, Sumit},
  journal={arXiv preprint arXiv:2305.13245},
  year={2023}
}

@article{dai2024deepseekmoe,
  title={DeepSeekMoE: Towards Ultimate Expert Specialization in Mixture-of-Experts Language Models},
  author={Dai, Damai and Deng, Chengqi and Zhao, Chenggang and Xu, RX and Gao, Huazuo and Chen, Deli and Li, Jiashi and Zeng, Wangding and Yu, Xingkai and Wu, Y and others},
  journal={arXiv e-prints},
  pages={arXiv--2401},
  year={2024}
}

@article{liu2024deepseekv2,
  title={Deepseek-v2: A strong, economical, and efficient mixture-of-experts language model},
  author={Liu, Aixin and Feng, Bei and Wang, Bin and Wang, Bingxuan and Liu, Bo and Zhao, Chenggang and Dengr, Chengqi and Ruan, Chong and Dai, Damai and Guo, Daya and others},
  journal={arXiv preprint arXiv:2405.04434},
  year={2024}
}

@article{yang2024qwen25,
  title={Qwen2.5 Technical Report},
  author={Qwen An Yang and Baosong Yang and Beichen Zhang and Binyuan Hui and Bo Zheng and Bowen Yu and Chengyuan Li and Dayiheng Liu and Fei Huang and Guanting Dong and Haoran Wei and Huan Lin and Jian Yang and Jianhong Tu and Jianwei Zhang and Jianxin Yang and Jiaxin Yang and Jingren Zhou and Junyang Lin and Kai Dang and Keming Lu and Keqin Bao and Kexin Yang and Le Yu and Mei Li and Mingfeng Xue and Pei Zhang and Qin Zhu and Rui Men and Runji Lin and Tianhao Li and Tingyu Xia and Xingzhang Ren and Xuancheng Ren and Yang Fan and Yang Su and Yi-Chao Zhang and Yunyang Wan and Yuqi Liu and Zeyu Cui and Zhenru Zhang and Zihan Qiu and Shanghaoran Quan and Zekun Wang},
  journal={arXiv preprint arXiv:2412.15115},
  year={2024}
}

@article{mmlu,
  title={Measuring massive multitask language understanding},
  author={Hendrycks, Dan and Burns, Collin and Basart, Steven and Zou, Andy and Mazeika, Mantas and Song, Dawn and Steinhardt, Jacob},
  journal={arXiv preprint arXiv:2009.03300},
  year={2020}
}

@article{mmlupro,
  author       = {Yubo Wang and
                  Xueguang Ma and
                  Ge Zhang and
                  Yuansheng Ni and
                  Abhranil Chandra and
                  Shiguang Guo and
                  Weiming Ren and
                  Aaran Arulraj and
                  Xuan He and
                  Ziyan Jiang and
                  Tianle Li and
                  Max Ku and
                  Kai Wang and
                  Alex Zhuang and
                  Rongqi Fan and
                  Xiang Yue and
                  Wenhu Chen},
  title        = {{MMLU-Pro}: {A} More Robust and Challenging Multi-Task Language Understanding
                  Benchmark},
  journal      = {CoRR},
  volume       = {abs/2406.01574},
  year         = {2024}
}

@inproceedings{bbh,
  author       = {Mirac Suzgun and
                  Nathan Scales and
                  Nathanael Sch{\"{a}}rli and
                  Sebastian Gehrmann and
                  Yi Tay and
                  Hyung Won Chung and
                  Aakanksha Chowdhery and
                  Quoc V. Le and
                  Ed H. Chi and
                  Denny Zhou and
                  Jason Wei},
  title        = {Challenging {BIG-Bench} Tasks and Whether Chain-of-Thought Can Solve
                  Them},
  booktitle    = {{ACL} (Findings)},
  pages        = {13003--13051},
  publisher    = {Association for Computational Linguistics},
  year         = {2023}
}

@article{supergpqa,
  title={{SuperGPQA}: Scaling {LLM} evaluation across 285 graduate disciplines},
  author={Du, Xinrun and Yao, Yifan and Ma, Kaijing and Wang, Bingli and Zheng, Tianyu and Zhu, King and Liu, Minghao and Liang, Yiming and Jin, Xiaolong and Wei, Zhenlin and others},
  journal={arXiv preprint arXiv:2502.14739},
  year={2025}
}

@article{gsm8k,
  author       = {Karl Cobbe and
                  Vineet Kosaraju and
                  Mohammad Bavarian and
                  Mark Chen and
                  Heewoo Jun and
                  Lukasz Kaiser and
                  Matthias Plappert and
                  Jerry Tworek and
                  Jacob Hilton and
                  Reiichiro Nakano and
                  Christopher Hesse and
                  John Schulman},
  title        = {Training Verifiers to Solve Math Word Problems},
  journal      = {CoRR},
  volume       = {abs/2110.14168},
  year         = {2021}
}

@inproceedings{math,
  author       = {Dan Hendrycks and
                  Collin Burns and
                  Saurav Kadavath and
                  Akul Arora and
                  Steven Basart and
                  Eric Tang and
                  Dawn Song and
                  Jacob Steinhardt},
  title        = {Measuring Mathematical Problem Solving With the {MATH} Dataset},
  booktitle    = {NeurIPS Datasets and Benchmarks},
  year         = {2021}
}

@inproceedings{hellaswag,
  author       = {Rowan Zellers and
                  Ari Holtzman and
                  Yonatan Bisk and
                  Ali Farhadi and
                  Yejin Choi},
  editor       = {Anna Korhonen and
                  David R. Traum and
                  Llu{\'{\i}}s M{\`{a}}rquez},
  title        = {{HellaSwag}: Can a Machine Really Finish Your Sentence?},
  booktitle    = {Proceedings of the 57th Conference of the Association for Computational
                  Linguistics, {ACL} 2019, Florence, Italy, July 28- August 2, 2019,
                  Volume 1: Long Papers},
  pages        = {4791--4800},
  publisher    = {Association for Computational Linguistics},
  year         = {2019},
  url          = {https://doi.org/10.18653/v1/p19-1472},
  doi          = {10.18653/v1/p19-1472},
  bibsource    = {dblp computer science bibliography, https://dblp.org}
}

@inproceedings{piqa,
  author       = {Yonatan Bisk and
                  Rowan Zellers and
                  Ronan Le Bras and
                  Jianfeng Gao and
                  Yejin Choi},
  title        = {{PIQA:} Reasoning about Physical Commonsense in Natural Language},
  booktitle    = {The Thirty-Fourth {AAAI} Conference on Artificial Intelligence, {AAAI}
                  2020, The Thirty-Second Innovative Applications of Artificial Intelligence
                  Conference, {IAAI} 2020, The Tenth {AAAI} Symposium on Educational
                  Advances in Artificial Intelligence, {EAAI} 2020, New York, NY, USA,
                  February 7-12, 2020},
  pages        = {7432--7439},
  publisher    = {{AAAI} Press},
  year         = {2020},
  url          = {https://doi.org/10.1609/aaai.v34i05.6239},
  doi          = {10.1609/aaai.v34i05.6239},
  bibsource    = {dblp computer science bibliography, https://dblp.org}
}

@article{arc,
  author       = {Peter Clark and
                  Isaac Cowhey and
                  Oren Etzioni and
                  Tushar Khot and
                  Ashish Sabharwal and
                  Carissa Schoenick and
                  Oyvind Tafjord},
  title        = {Think you have Solved Question Answering? {Try} {ARC}, the {AI2} Reasoning
                  Challenge},
  journal      = {CoRR},
  volume       = {abs/1803.05457},
  year         = {2018}
}

@article{humaneval,
  author       = {Mark Chen and
                  Jerry Tworek and
                  Heewoo Jun and
                  Qiming Yuan and
                  Henrique Pond{\'{e}} de Oliveira Pinto and
                  Jared Kaplan and
                  Harrison Edwards and
                  Yuri Burda and
                  Nicholas Joseph and
                  Greg Brockman and
                  Alex Ray and
                  Raul Puri and
                  Gretchen Krueger and
                  Michael Petrov and
                  Heidy Khlaaf and
                  Girish Sastry and
                  Pamela Mishkin and
                  Brooke Chan and
                  Scott Gray and
                  Nick Ryder and
                  Mikhail Pavlov and
                  Alethea Power and
                  Lukasz Kaiser and
                  Mohammad Bavarian and
                  Clemens Winter and
                  Philippe Tillet and
                  Felipe Petroski Such and
                  Dave Cummings and
                  Matthias Plappert and
                  Fotios Chantzis and
                  Elizabeth Barnes and
                  Ariel Herbert{-}Voss and
                  William Hebgen Guss and
                  Alex Nichol and
                  Alex Paino and
                  Nikolas Tezak and
                  Jie Tang and
                  Igor Babuschkin and
                  Suchir Balaji and
                  Shantanu Jain and
                  William Saunders and
                  Christopher Hesse and
                  Andrew N. Carr and
                  Jan Leike and
                  Joshua Achiam and
                  Vedant Misra and
                  Evan Morikawa and
                  Alec Radford and
                  Matthew Knight and
                  Miles Brundage and
                  Mira Murati and
                  Katie Mayer and
                  Peter Welinder and
                  Bob McGrew and
                  Dario Amodei and
                  Sam McCandlish and
                  Ilya Sutskever and
                  Wojciech Zaremba},
  title        = {Evaluating Large Language Models Trained on Code},
  journal      = {CoRR},
  volume       = {abs/2107.03374},
  year         = {2021}
}

@article{qwen2.5,
  title={Qwen2. 5 technical report},
  author={Yang, An and Yang, Baosong and Zhang, Beichen and Hui, Binyuan and Zheng, Bo and Yu, Bowen and Li, Chengyuan and Liu, Dayiheng and Huang, Fei and Wei, Haoran and others},
  journal={arXiv preprint arXiv:2412.15115},
  year={2024}
}

@inproceedings{rein2024gpqa,
  title={Gpqa: A graduate-level google-proof q\&a benchmark},
  author={Rein, David and Hou, Betty Li and Stickland, Asa Cooper and Petty, Jackson and Pang, Richard Yuanzhe and Dirani, Julien and Michael, Julian and Bowman, Samuel R},
  booktitle={First Conference on Language Modeling},
  year={2024}
}

@article{sakaguchi2021winogrande,
  title={Winogrande: An adversarial winograd schema challenge at scale},
  author={Sakaguchi, Keisuke and Bras, Ronan Le and Bhagavatula, Chandra and Choi, Yejin},
  journal={Communications of the ACM},
  volume={64},
  number={9},
  pages={99--106},
  year={2021},
  publisher={ACM New York, NY, USA}
}

@inproceedings{xiong2024superbench,
author = {Xiong, Yifan and Jiang, Yuting and Yang, Ziyue and Qu, Lei and Zhao, Guoshuai and Liu, Shuguang and Zhong, Dong and Pinzur, Boris and Zhang, Jie and Wang, Yang and Jose, Jithin and Pourreza, Hossein and Baxter, Jeff and Datta, Kushal and Ram, Prabhat and Melton, Luke and Chau, Joe and Cheng, Peng and Xiong, Yongqiang and Zhou, Lidong},
title = {SuperBench: Improving Cloud AI Infrastructure Reliability with Proactive Validation},
booktitle = {USENIX ATC},
year = {2024},
month = {July},
publisher = {USENIX Association},
pages = {835-850},
}

@misc{microsoft2018openpai,  
  author       = {Microsoft},  
  title        = {Open Platform for AI},  
  year         = {2018},  
  howpublished = {\url{https://github.com/microsoft/pai}}
}

@article{shoeybi2019megatron,
  title={Megatron-lm: Training multi-billion parameter language models using model parallelism},
  author={Shoeybi, Mohammad and Patwary, Mostofa and Puri, Raul and LeGresley, Patrick and Casper, Jared and Catanzaro, Bryan},
  journal={arXiv preprint arXiv:1909.08053},
  year={2019}
}
